# Spatial localization of magnetic-branch force contrasts in substitutional Mn-doped Ge: A first-principles study

**ChienHsu Chen**

Accelerator Laboratory, Nuclear Science and Technology Development Center, National Tsing Hua University, Hsinchu, Taiwan

Corresponding author: achchen@mx.nthu.edu.tw

## Abstract

Magnetic energy gaps do not reveal the spatial distribution of magnetic-state-dependent atomic forces. We compare ferromagnetic and antiferromagnetic forces at identical nuclear positions in 64-site substitutional Mn-doped Ge using collinear PBE calculations. For nearest-neighbour Mn pairs, the radial force projection reverses sign between magnetic branches. In the second-neighbour geometry, the shared Ge neighbour has the largest force contrast, about 2.2 times that of equally near Ge sites coordinated to one Mn. Bridge-displacement gap slopes agree with direct force derivatives across four numerical settings. Tightening the electronic threshold by 100-fold preserves magnetic ordering and the leading site, while the force-contrast vector changes by 1.08% despite a nearly unchanged energy gap. The results identify coordination-associated localization of magnetic-branch force contrasts within the frozen model, without establishing relaxed distortions, full numerical convergence or a unique exchange mechanism.

## 1. Introduction

Transition-metal impurities in group-IV semiconductors couple local magnetic moments to a directional covalent host. In substitutional Mn-doped Ge, the energetic preference between magnetic configurations depends on impurity separation, geometry and electronic environment [1–4]. Earlier studies examined oscillatory magnetic coupling and electronic codoping [2,3]. Picozzi and Ležaić discussed the dependence of second-neighbour [110] exchange on the intervening atomic environment [4]. Induced polarization of neighbouring Ge is also established [3]. The present study does not claim a new magnetic ordering or the first observation of bridge-site polarization.

The question addressed here concerns a different observable: the full atomic-force contrast between self-consistent magnetic branches at unchanged nuclei. A magnetic energy gap specifies the relative variational energy of two sampled branches at one geometry. It does not specify the gradient of that gap over the nuclear coordinates. Mapping this gradient makes the spatial distribution of the magnetic response directly accessible, without combining magnetic and ionic relaxation.

The second-neighbour Mn geometry offers a useful local comparison. One Ge atom belongs to the first-neighbour shells of both Mn impurities. Other Ge sites have the same distance to their closest Mn but belong to only one Mn first-neighbour shell. These sites are therefore not equivalent by coordination, despite their equal nearest-impurity distance. Comparing their complete force vectors tests whether the magnetic response is associated with shared coordination. Moving the shared atom provides a separate energy-based check of the corresponding local gradient.

Magnetic degrees of freedom are already recognized in atomistic modelling. Neural-network descriptions of Mn–Ge compounds have been reported [5], while spin-aware neural-network and magnetic moment tensor potentials distinguish magnetic states [6,7]. We do not train or test these potentials. A conditional force-target ambiguity is included only as an algebraic consequence of the sampled multi-branch data.

The electronic-structure approximation remains important. PBE, Hubbard-corrected and hybrid-functional calculations can give different descriptions of Mn-related states in Ge [8]. Finite supercells, reciprocal-space sampling and electronic smearing also affect numerical results. We therefore distinguish reproducibility within a frozen PBE model from an equilibrium structural prediction or a uniquely identified exchange mechanism.

We investigate three questions: whether the projected Mn–Mn radial force changes sign between magnetic branches; whether a shared Ge site dominates the force contrast; and whether its explicit displacement reproduces the gap gradient inferred from forces. Eight reference calculations and a complete 24-case numerical-sensitivity campaign supply 32 final outputs and 16 matched magnetic pairs. The central result is a site-resolved pattern associated with shared-shell coordination, not a general magnetic phase diagram.

## 2. Computational methods

### 2.1. Frozen structures and magnetic branches

The reference geometry is a cubic 64-site diamond-Ge supercell with prescribed side length 11.485532 Å. Replacing two Ge atoms by Mn gives $Ge_{62}Mn_2$, or 3.125% Mn. In the first-neighbour (1NN) pair, original sites #1 and #5 are Mn, separated by 2.487 Å. In the second-neighbour (2NN) pair, Mn occupies sites #1 and #2, separated by 4.061 Å. Site #5 then remains Ge and is a first neighbour of both Mn atoms. Each Mn–Ge#5 distance is 2.487 Å. Ge#6 and Ge#30 have this same closest-Mn distance but are first neighbours of only one Mn (Fig. 1).

The FM and AFM calculations in each pair use exactly the same cell and nuclear coordinates. A displacement series moves only Ge#5 along Cartesian $x$ by $q = -0.05$, 0 and +0.05 Å; all other nuclei remain fixed. The zero-displacement calculations reuse the unshifted 2NN pair. Thus each numerical setting contains eight distinct cases: 1NN FM/AFM, 2NN FM/AFM, and four displaced endpoints. Original atomic ordering is preserved. Site #5 is Mn in 1NN and Ge in 2NN; these labels are not interchangeable.

Every compared cell is neutral, with 898 electrons in the archived Quantum ESPRESSO XML. Magnetic branches are not different net-charge states. The cell size is a frozen modelling constraint. Its independent pristine-Ge relaxation provenance is not established by this dataset. No ionic relaxation or equilibrium defect structure is claimed.

### 2.2. Electronic-structure settings and evidence scope

Calculations use Quantum ESPRESSO pw.x 7.5 [9,10], collinear spin-polarized PBE [11], and scalar-relativistic ultrasoft pseudopotentials from PSLibrary [13]. The exact files are:

```
Ge.pbe-dn-rrkjus_psl.1.0.0.UPF
Mn.pbe-spn-rrkjus_psl.0.3.1.UPF
```

Their complete hashes and retained bytes are supplied with the data. Reference settings are 60/480 Ry wavefunction/density cutoffs, an unshifted Γ-centred $3 \times 3 \times 3$ mesh, and Marzari–Vanderbilt cold smearing of 0.01 Ry [12]. The electronic convergence threshold is $10^{-8}$ Ry. Collinear `nspin=2` is used without spin–orbit coupling. Mn starting-magnetization seeds of ±0.5 initialize orientations, not constrained final local moments.

The numerical campaign changes one specified input at a time. Arm A uses a $4 \times 4 \times 4$ mesh, arm B uses 70/560 Ry cutoffs, and arm C uses a $10^{-10}$ Ry SCF threshold. All remaining physical input fields match the corresponding reference case. Each arm contains eight electronically converged cases. No ionic-force stopping criterion applies because the nuclei are not minimized.

The immutable v9 archive contains eight reference and 21 sensitivity outputs. Three separately archived post-v9 terminals—C03, C04 and C05—complete the sensitivity set to 24/24 and the scientific dataset to 32 distinct final XML files. They are used here with their separate provenance; v9 was not repacked. This is the complete dataset for the present two-Mn paper, not a merger with the separate three-Mn T123 or 32-site single-Mn studies.

Raw output integrity, final electronic convergence and input equivalence were checked separately from physical validity. The analysis retains all 2,048 site-force and atomic-sphere-moment records. Decimal differences of XML energy strings are independently checked with rational arithmetic. The original scientific inputs, stdout, stderr and XML are preserved byte-for-byte. Terminal warnings remain in the dataset. Host execution differs for some cases: in particular, cutoff case B07, the positive-displacement FM endpoint, ran on Dell R730, while its AFM companion B08 ran on Dell T7810. This is not a strict identical-runtime comparison, despite matching nominal input physics. Archived runs use 4, 6 or 22 MPI processes with one thread per process; the per-case layout is retained in cases.csv. The numerical arms are not separate tests of parallel-reduction or binary-environment reproducibility.

Derived physical quantities are usually reported to two or three significant figures in the narrative and three or four in main-text tables. Extra digits are retained only for close numerical comparisons, including derivatives and cosines near unity. All differences, ratios and plotted coordinates are calculated from unrounded values before display rounding. SI tables retain diagnostic precision, while the original XML and full-precision CSV values are unchanged. Reported digits do not establish physical accuracy or numerical convergence, and no generic DFT uncertainty is assigned. Prescribed inputs, exact counts, unit-conversion constants and file identifiers retain the precision needed for reproducibility. Unit conversion uses 1 Ry = 13.605693122994 eV and 1 bohr = 0.529177210903 Å; XML forces are in Hartree/bohr, not Ry/bohr.

## 2.3. Energies, atomic forces and the radial coordinate

Let $F_b(\mathbf{R})$ be the converged variational energy of magnetic branch $b$. For cold smearing, the QE variational total is used consistently with the forces; the separately printed internal energy is not substituted. Define

$$\Delta F = F_{\mathrm{AFM}} - F_{\mathrm{FM}}, \qquad \mathbf{f}_i^b = -\frac{\partial F_b}{\partial \mathbf{R}_i} \tag{1}$$

For Mn sites A and B, $\hat{\mathbf{r}}_{AB}$ points along the unique shortest periodic displacement from A to B. The radial generalized force is

$$Q_b = \frac{(\mathbf{f}_B^b - \mathbf{f}_A^b)\cdot\hat{\mathbf{r}}_{AB}}{2} \tag{2}$$

A symmetric outward displacement of the two Mn atoms by a total separation increment $dr$ gives $dF_b/dr = -Q_b$, and therefore $d\Delta F/dr = Q_{\mathrm{FM}} - Q_{\mathrm{AFM}}$. Positive $Q_b$ denotes an outward relative projection. Equation (2) uses total site forces from the entire environment, not an isolated Mn–Mn pair force.

## 2.4. Spatial force-contrast metrics

At identical nuclei, define

$$\Delta\mathbf{f}_i = \mathbf{f}_i^{\mathrm{AFM}} - \mathbf{f}_i^{\mathrm{FM}}, \qquad S_i = \frac{\|\Delta\mathbf{f}_i\|^2}{\sum_j \|\Delta\mathbf{f}_j\|^2} \tag{3}$$

The quantity $S_i$ is a squared-force-contrast fraction, not an exchange-energy or bonding-energy fraction. Coordination is determined from periodic minimum-image distances to both Mn atoms. The 64 force-contrast vectors are concatenated in unchanged atomic order into a 192-component vector $\mathbf{D}$. Relative to the matching reference $\mathbf{D}_0$, we report

$$c = \frac{\mathbf{D}\cdot\mathbf{D}_0}{\|\mathbf{D}\|\|\mathbf{D}_0\|}, \qquad \eta = 100\,\frac{\|\mathbf{D}-\mathbf{D}_0\|}{\|\mathbf{D}_0\|} \tag{4}$$

Cosine similarity describes the complete signed-vector pattern. The relative difference also detects amplitude changes. We repeat the cosine after excluding the three Ge#5 components. Neither statistic alone establishes asymptotic convergence.

### 2.5. Bridge-displacement derivative check

For the 2NN cell, $\mathbf{R}_5(q) = \mathbf{R}_5(0) + q\hat{\mathbf{x}}$, with $h = 0.05$ Å. Both magnetic branches use identical nuclei at every sampled $q$. Define

$$D_h = \frac{\Delta F(+h) - \Delta F(-h)}{2h}, \qquad D_f = -\Delta f_{5x}(0) \tag{5}$$

The reported relative discrepancy is $100\left|D_h - D_f\right|/\left|D_f\right|$. This denominator is used consistently in the text, figures and tables. The signed Cartesian component $\Delta f_{5x}$, not the vector norm, enters Eq. (5). A comparison with the mean of the endpoint force derivatives is retained separately in the dataset. These finite-interval tests do not separate displacement truncation, electronic sampling and model errors.

### 2.6. Electronic markers and reproducible analysis

Final magnetic labels are checked against the signs of Mn atomic-sphere spin projections. These partition-dependent moments are not whole-space atomic magnetizations or orbital-resolved Mn occupations. Fermi energies are auxiliary branch markers, not a mechanism decomposition. Stable spin signs under displacement cannot exclude every metastable electronic solution.

For two equally weighted force targets at identical nuclei, their unconstrained common least-squares target is the arithmetic mean. The resulting conditional per-atom-vector RMS is $\| \mathbf{D} \| / \left(2\sqrt{N}\right)$, where $N = 64$. This algebraic statistic is not a measured error of any fitted potential or of a selected ground-state surface.

The complete numerical analysis is implemented in the accompanying Python scripts; AI assistance is described in the dedicated declaration. The executable scripts, package versions, file hashes and tabulated arithmetic define the reproducible analysis. Figures are deterministic Matplotlib visualizations of the retained numerical data, not generatively synthesized research images.

## 3. Results

### 3.1. Magnetic ordering and radial projections

The reference AFM-minus-FM gap is approximately −569 meV/cell for 1NN and +132 meV/cell for 2NN. Thus AFM is lower among the sampled 1NN branches and FM among the sampled unshifted 2NN branches. A denser mesh preserves these signs but changes the magnitudes to approximately −551 and +141 meV/cell. The corresponding absolute relative changes are 3.2% and 7.1%. The cutoff and tight-SCF changes are much smaller (Table 1); this does not establish physical or force convergence.

**Table 1.** AFM-minus-FM variational gaps (meV per 64-site cell). Reference, mesh and cutoff columns originate in v9. The 2NN tight-SCF entry uses separately archived C03/C04. Display values are rounded to 0.1 meV. Identical rounded entries do not imply identical source energies; the finer differences are retained in Table S2 and discussed separately.

| Geometry | Reference | $4^3$ mesh | 70/560 Ry | Tight SCF |
|---|---|---|---|---|
| 1NN | -568.8 | -550.9 | -568.8 | -568.8 |
| 2NN | +132.0 | +141.3 | +132.0 | +132.0 |

The radial projection addresses a different observable (Fig. 2). For unshifted 1NN, the reference values are approximately +0.436 eV/Å (FM) and −0.510 eV/Å (AFM), so the sign reverses. Both unshifted 2NN branches project inward, at approximately −0.0755 and −0.102 eV/Å. These signs persist across the tested numerical settings. They do not specify relaxed Mn–Mn separations.

**Table 2.** Radial generalized forces at the unshifted pair geometries (eV/Å). Positive values mean outward relative separation. Values refer to total site forces, not a two-body force. Columns are rounded independently after calculation; additional digits are retained in Table S3.

| Setting | Geometry | $Q_{\mathrm{FM}}$ | $Q_{\mathrm{AFM}}$ | $Q_{\mathrm{FM}} - Q_{\mathrm{AFM}}$ |
|---|---|---|---|---|
| Reference | 1NN | +0.4357 | -0.5103 | +0.9460 |
| Reference | 2NN | -0.0755 | -0.1016 | +0.0261 |
| $4^3$ mesh | 1NN | +0.4544 | -0.4998 | +0.9543 |
| $4^3$ mesh | 2NN | -0.0748 | -0.0987 | +0.0239 |
| 70/560 Ry | 1NN | +0.4382 | -0.5048 | +0.9430 |
| 70/560 Ry | 2NN | -0.0753 | -0.1012 | +0.0259 |
| Tight SCF | 1NN | +0.4352 | -0.5104 | +0.9456 |
| Tight SCF | 2NN | -0.0752 | -0.1007 | +0.0255 |

The tight-SCF 2NN pair changes the radial gap gradient from +0.0261 to +0.0255 eV/Å, a decrease of approximately 2.19%. The energy gap changes by only about $1.23\times10^{-5}$%. This difference directly shows why a stable magnetic gap cannot serve as a uniform bound on force or gap-gradient sensitivity.

### 3.2. The shared Ge site dominates the local contrast

Ge#5 lies 2.487 Å from each Mn and forms a Mn–Ge–Mn angle of 109.5°. Ge#6 and Ge#30 have the same closest-Mn distance but are 6.258 Å from the other Mn. Their geometrically defined angle to that remote Mn is not a two-bond bridge angle. The controlled distinction is shared versus single-Mn first-neighbour coordination, not merely the shortest impurity distance.

The shared Ge has the largest full-vector force contrast among all 64 sites at the unshifted 2NN geometry (Fig. 3). Reference norms are 0.348 eV/Å for Ge#5, 0.157 eV/Å for Ge#6 and 0.155 eV/Å for Ge#30. The shared-site contrast is therefore about 2.23–2.25 times that of these equally near single-Mn neighbours. At the denser mesh the corresponding norms are 0.332, 0.141 and 0.142 eV/Å. Ge#5 remains first under the cutoff and tight-SCF changes (Table 3).

Its squared-contrast share is 36.8% at reference, 37.5% on the denser mesh and approximately 36.8% for both remaining arms. This share is not a fraction of magnetic exchange energy. The site-resolved rank and equal-distance comparison are more discriminating than the total share of the Ge subsystem, which contains 62 of the 64 atoms. The observation concerns this sampled environment; it is not a universal statement about all GeMn defect topologies.

### 3.3. Spatial resemblance does not imply amplitude convergence

The reference and denser-mesh 192-component force-contrast vectors have cosine 0.99451. After removing Ge#5, the cosine remains 0.99135. Thus their resemblance is not due only to the largest atom. Nevertheless, their relative vector difference is 11.6%. The cutoff comparison gives cosine 0.99982 and a 1.88% vector difference. A high cosine and a stable rank therefore coexist with non-negligible changes in force components.

**Table 3.** Unshifted 2NN force-contrast metrics. Norms are in eV/Å; shares and vector changes are percentages. Cosines compare all signed components with the reference; the exclusion column removes Ge#5. C03/C04 are separately

archived post-v9 data. Norms have four significant figures; cosines retain five decimal places to resolve departures from unity. Full diagnostic values are in Tables S5 and S6.

| Setting | Ge#5 norm | Share (%) | Cosine | Excl. #5 | Change (%) |
|---|---|---|---|---|---|
| Reference | 0.3484 | 36.8 | 1.00000 | 1.00000 | 0 |
| $4^3$ mesh | 0.3322 | 37.5 | 0.99451 | 0.99135 | 11.6 |
| 70/560 Ry | 0.3486 | 36.8 | 0.99982 | 0.99977 | 1.88 |
| Tight SCF | 0.3489 | 36.8 | 0.99994 | 0.99992 | 1.08 |

Tightening the SCF threshold by 100-fold decreases the 2NN gap by approximately $1.63\times10^{-5}$ meV/cell; both gaps round to +132 meV/cell. The small difference is a within-model SCF diagnostic, not a physical error bound. The corresponding force-contrast cosine is 0.99994, but the vector changes by 1.08%. The bridge norm increases by about 0.126%, and its rank remains 1. Meanwhile, the radial gap gradient changes by 2.19%. These are three different force diagnostics; the stability of one cannot replace the others. Figure 6 compares their sensitivities on the same scale.

### 3.4. Explicit bridge displacement checks the gap gradient

For all four numerical settings, the magnetic gap decreases monotonically as Ge#5 moves from −0.05 to +0.05 Å (Fig. 4). The reference gaps round to 149, 132, and 115 meV/cell, respectively. The tight-SCF gaps have the same values at this reporting precision. The full set of gaps is listed in Table S2; no missing endpoint is interpolated or imputed.

**Table 4.** Bridge-gap derivatives at h = 0.05 Å. Both derivatives and their absolute difference are in eV/Å. Relative discrepancy is $100|D_h - D_f|/|D_f|$ and is an internal consistency comparison, not a physical error bar. Derivatives retain five decimal places solely to resolve the comparison. Differences and percentages use unrounded source values, not the rounded table entries; Table S7 provides additional diagnostic digits.

| Setting | $D_h$ | $D_f$ | Abs. difference | Difference (%) |
|---|---|---|---|---|
| Reference | -0.34835 | -0.34844 | 0.0000929 | 0.0267 |
| $4^3$ mesh | -0.33394 | -0.33221 | 0.00174 | 0.523 |
| 70/560 Ry | -0.34841 | -0.34861 | 0.000199 | 0.0571 |
| Tight SCF | -0.34835 | -0.34889 | 0.000539 | 0.155 |

Across the four settings, energy-derived slopes range from approximately −0.348 to −0.334 eV/Å and agree in sign with the corresponding direct-force derivatives. Relative discrepancies are 0.0267%, 0.523%, 0.0571% and 0.155% (Fig. 5). The tighter electronic threshold does not make the finite-step discrepancy vanish. These differences combine finite-displacement and numerical effects; they are not estimates of the uncertainty of PBE.

Moving the bridge changes both Mn–Ge distances and the intervening angle. Consequently, the observed derivative cannot be assigned uniquely to bond stretching or angle bending. The Mn atoms remain fixed throughout. The individual AFM Ge#5 force also changes strongly with the mesh even while the branch contrast remains near 0.33–0.35 eV/Å. Convergence of a force difference is therefore distinct from convergence of either branch’s equilibrium position.

### 3.5. Final magnetic markers and the completed negative endpoint

For reference 2NN, the final Mn sphere moments are approximately +3.34/+3.34 μB in FM and +3.39/−3.39 μB in AFM. The shared Ge moment is approximately −0.124 μB in FM and close to zero in AFM. These signs persist under the tested displacement and numerical changes. The moments support the branch labels but do not specify a unique electronic mechanism.

C05 closes the tight-SCF negative-displacement FM/AFM pair with C06. Its final XML records 24 iterations and an estimated SCF error of $8.16\times10^{-11}$ Ry, below the requested $10^{-10}$ Ry. The negative-endpoint gap is approximately +149 meV/cell. Ge#5 remains the largest contrast at approximately 0.345 eV/Å and accounts for 37.9% of the squared contrast. The full field differs from its matching reference by 0.871%, whereas the absolute relative gap change is $1.85\times10^{-5}$% (Table S11). Separately, the tight-SCF three-point check has a 0.155% discrepancy between the energy-derived slope and the direct-force derivative (Table 4); this is not a slope change relative to reference. This completed endpoint enables the fourth three-point check; it is not a new physical geometry beyond the planned campaign.

Reference Fermi energies lie near 7.6 eV; printed branch-resolved values are retained in Section S6 and cases.csv. They are electronic-branch markers, not orbital occupations. In particular, the approximately −0.569 eV 1NN gap is not automatically an isolated Heisenberg exchange constant.

## 4. Discussion

### 4.1. Coordination-associated localization

The observed energy ordering is consistent with the broader dependence of Mn–Mn coupling on local environment documented in earlier work [1–4]. The contribution here is the matched, same-nuclei comparison of every atomic-force vector and its association with coordination. The shared Ge response exceeds those of equally near single-Mn neighbours. Its leading rank persists across all four numerical settings and at the sampled displacement endpoints.

The full signed-vector comparison strengthens this observation beyond a single force component or an element-averaged statistic. Excluding the shared site still leaves a similar spatial pattern. The explicit displacement relates the local force contrast to a derivative of the magnetic gap. Together these results identify a reproducible organization of the gradient within the specified model. They do not establish a unique p–d hybridization, superexchange or other microscopic pathway.

### 4.2. Distinct convergence questions

The tight-SCF pair provides a particularly clear distinction between energy and force diagnostics: a gap change of roughly $10^{-5}$% accompanies a 1.08% force-vector change and a 2.19% radial-gradient change. This is a numerical sensitivity observation, not a general quantitative rule. The denser mesh produces the largest tested 2NN changes, about 7.1% in gap and 11.6% in full-vector contrast. Hence the present dataset does not establish complete k-point convergence of magnitudes.

The four bridge-gradient comparisons test the derivative convention and consistency of retained energies and forces at a finite displacement step. Their agreement does not establish step-size convergence, because only one nonzero step magnitude was used. Nor does it separate finite-step errors from changes in self-consistent occupations. Preserved Mn spin signs help track the intended branches but do not prove analytic continuity of the entire electronic state between endpoints.

The cutoff series has an additional runtime qualification: B07, the positive-displacement FM result, was produced on R730; B08 and the other cutoff cases were produced on T7810. Nominal scientific inputs match, but the series should not be described as an identical-environment repetition. The archived c_bands and IEEE warning lines remain available alongside the final convergence evidence. Their retention prevents normal termination from erasing negative numerical provenance.

### 4.3. Scope and consequences for force descriptions

C03/C04/C05 were not members of the immutable v9 archive. They are separately archived post-v9 terminals used here with explicit source mapping. The combined dataset now covers all 24 planned sensitivity cases. Source completeness does not resolve untested supercell sizes, smearing widths, exchange–correlation choices, noncollinear states, spin–orbit coupling or ionic relaxation. It also does not establish experimental phase stability or a Curie temperature.

A geometry-only single-valued prediction cannot match two unequal branch-force labels at the same nuclei without a branch descriptor or a branch-selection rule. However, modelling a selected ground-state surface is a different problem. The equal-weight conditional RMS values, 0.0862 eV/Å for reference 1NN and 0.0359 eV/Å for reference 2NN, are algebraic statistics of these labels. They are not empirical error floors of published potentials [5–7]. The present data instead suggest specific targets for a future spin-conditioned description: radial projections, full spatial contrasts, bridge-coordinate gap slopes and their numerical sensitivity.

## 5. Conclusions

At identical frozen nuclei, the sampled magnetic branches of substitutional Mn-doped Ge produce different site-force fields. The 1NN radial projection changes sign between FM and AFM, whereas both unshifted 2NN projections point inward. In the 2NN cell, the shared first-neighbour Ge atom has the largest contrast among all 64 sites. It exceeds equally near, d coordinated Ge sites by about a factor of 2.2 and retains its rank under the tested mesh, cutoff and SCF changes.

The complete 32-case dataset supplies four bridge-displacement checks. Energy-gap slopes and direct-force derivatives agree in sign, with sub-percent internal discrepancies under the stated convention. The tight-SCF comparison simultaneously exposes a near-constant magnetic gap and finite changes in the full force contrast and radial gap gradient. Thus energy-gap stability is not a substitute for atom-resolved force convergence.

These results describe coordination-associated localization within a finite, collinear PBE model. They do not establish relaxed lattice distortions, a unique exchange mechanism or a transferable interatomic potential. The retained raw outputs and executable analysis make both the supported pattern and its numerical limits inspectable.

## Data availability

The inputs, final XML, stdout/stderr, pseudopotential identities, complete site-force and moment tables, and analysis and plotting scripts are supplied as the accompanying GeMn Paper 1 research dataset. It contains 32 unique terminal cases: the 29 cases in the unchanged v9 archive plus separately archived C03/C04/C05. Source-object checksums and file-level lineage are included. The institutional custody objects require permission and are not public persistent repository records. No public DOI has yet been assigned. The accompanying dataset is available from the author for editorial and reviewer access.

## Funding

This research received no external funding.

## Declaration of Generative AI and AI-assisted Technologies

During the preparation of this work, the author used OpenAI ChatGPT and Codex to assist with scientific discussions, development and review of analysis scripts, computational data analysis, manuscript organization, and language editing. The numerical results and data visualizations were derived from archived Quantum ESPRESSO outputs by the documented scripts. The author retains responsibility for reviewing the AI-assisted outputs against the underlying computational evidence and for all scientific interpretations, conclusions, and content of this publication.

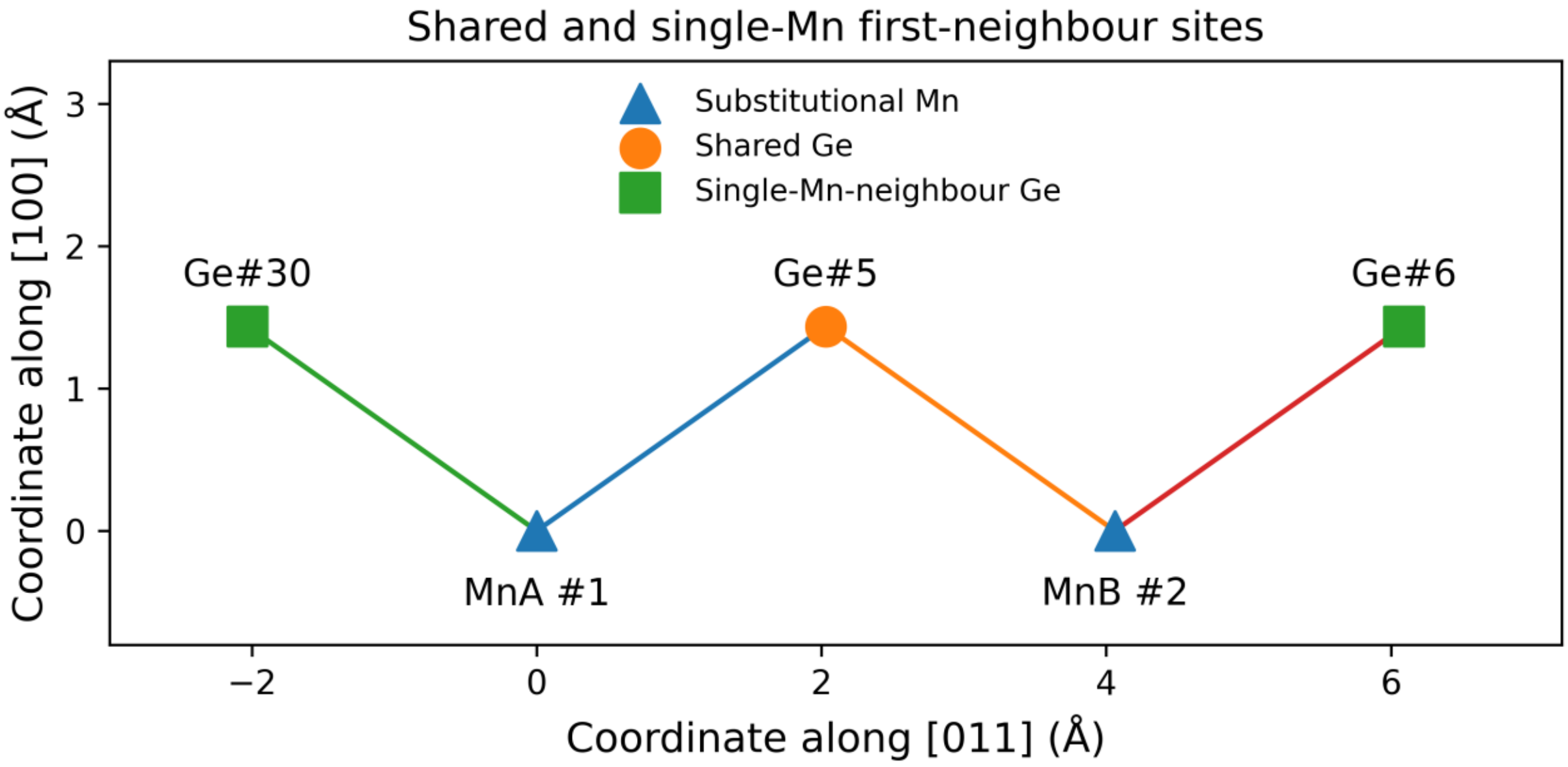

**Figure 1.** Local geometry reconstructed from the original frozen 2NN coordinates. Solid segments connect first-neighbour Mn–Ge sites and have length 2.487 Å. The Mn–Mn separation is 4.061 Å. Ge#5 is shared by both Mn shells, whereas Ge#6 and Ge#30 have only one first-neighbour Mn. Coordinates are projected onto [011] and [100]; this is a geometric construction, not an identified exchange pathway or a relaxed structure. Source: atoms.csv and coordination.csv.

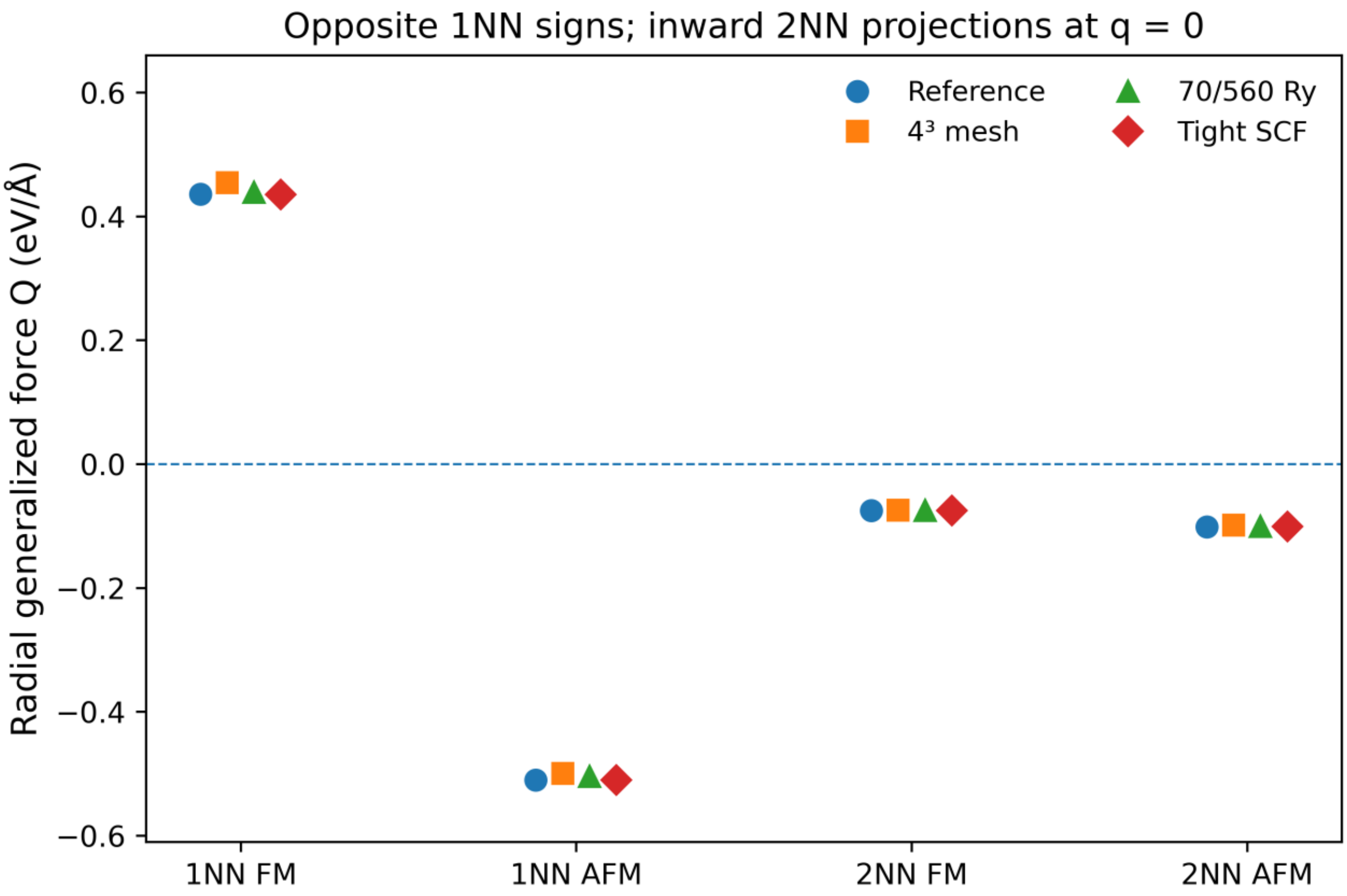

**Figure 2.** Radial total-site-force projections for the unshifted 1NN and 2NN geometries. Reference 1NN values are +0.436 and −0.510 eV/Å for FM and AFM; both 2NN values are negative. All four numerical settings preserve these signs. Horizontal marker offsets only prevent overlap and are not additional geometries. Positive Q means an outward relative projection, not an isolated pairwise force. Source: pairs.csv.

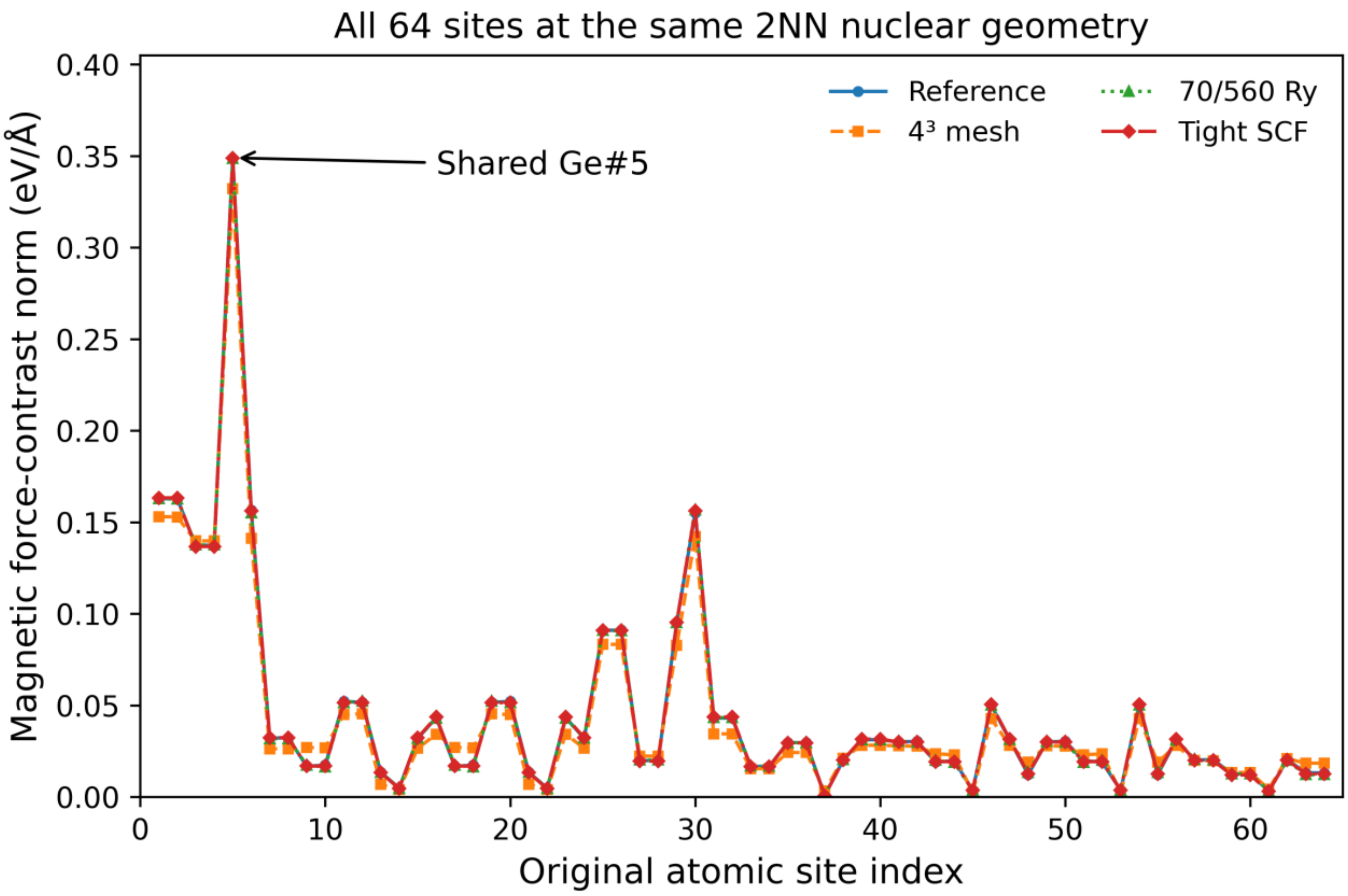

**Figure 3.** AFM-minus-FM force-contrast norms for all 64 sites in unchanged atomic order at the unshifted 2NN geometry. Ge#5 remains the maximum: approximately 0.348 eV/Å at reference and 0.349 eV/Å under tight SCF. Connecting lines guide site indices, not interatomic distance or chemical bonds. All settings use fixed nuclei; C03/C04 are post-v9 terminals. Source: force_contrasts.csv.

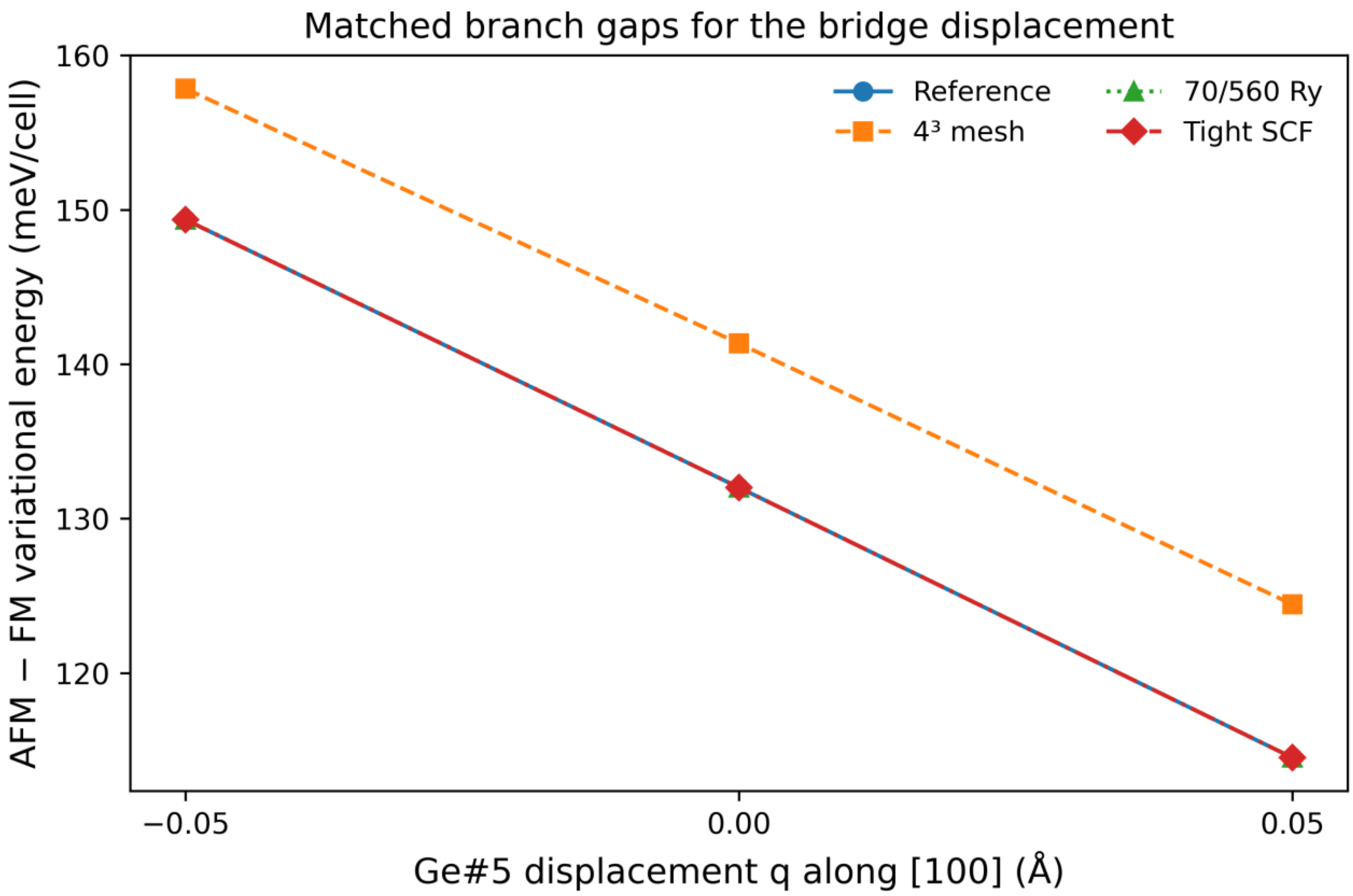

**Figure 4.** Magnetic variational gaps for Ge#5 displacements of −0.05, 0 and +0.05 Å, keeping the other 63 nuclei fixed. The reference gap decreases from approximately 149 to 115 meV/cell. Each point uses a complete FM/AFM pair. Lines are visual guides, not a fitted potential. The tight-SCF negative endpoint uses C05/C06; the cutoff positive endpoint combines R730 B07 with T7810 B08. Source: derivatives.csv.

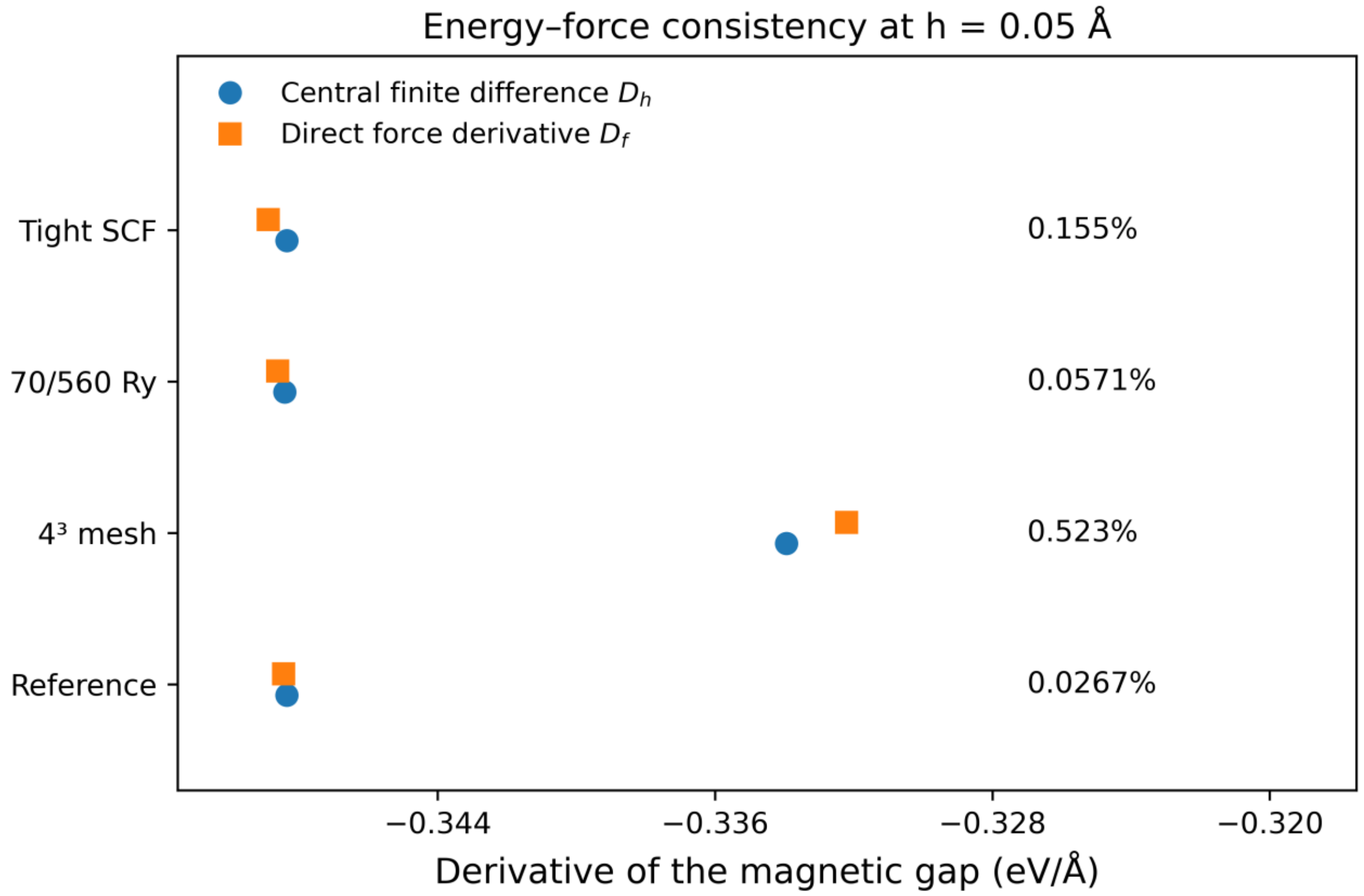

**Figure 5.** Central finite-difference slopes and direct Cartesian-force derivatives of the magnetic gap at h = 0.05 Å. Labels give $100|D_h - D_f|/|D_f|$: 0.0267%, 0.523%, 0.0571% and 0.155% for reference, mesh, cutoff and tight SCF. Small vertical marker offsets aid visibility. These are finite-interval internal checks, not physical uncertainty bars. Source: derivatives.csv.

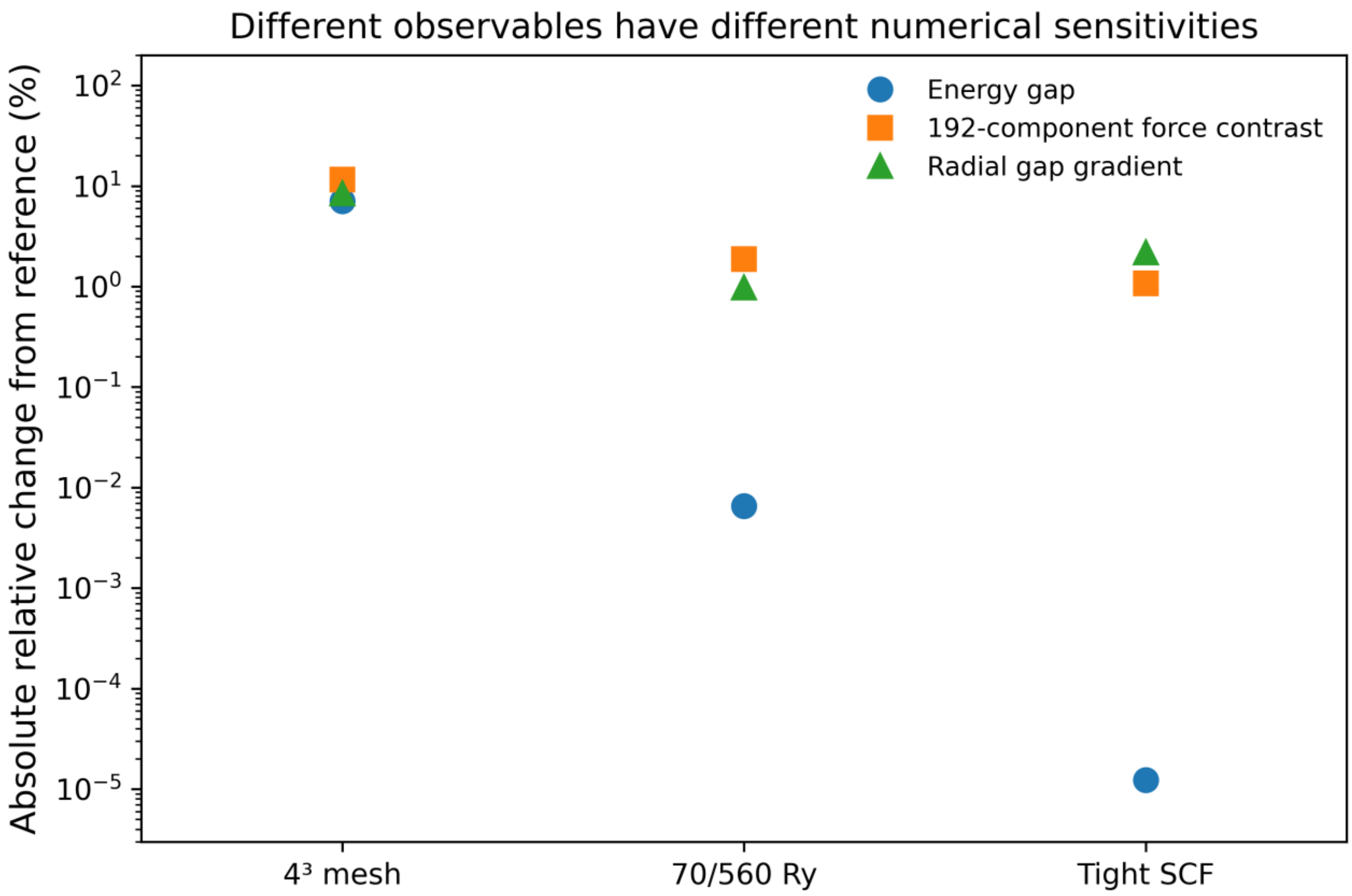

**Figure 6.** Absolute relative changes in three distinct observables at the same unshifted 2NN geometry: magnetic gap, complete 192-component force contrast and radial gap gradient. Each uses its own matching reference denominator. Under tight SCF the energy gap changes by approximately $1.23\times10^{-5}$%, the vector by 1.08%, and the radial gradient by 2.19%. The logarithmic axis exposes their separation; no common uncertainty scale is implied. Source: pairs.csv.